\documentclass[a4paper, 12pt]{article}
\usepackage{amsmath,amsfonts,amssymb,epsfig,subfigure}
\usepackage{amsmath,amsfonts,amssymb}

\newcommand{\be}{\begin{equation}}
\newcommand{\en}{\end{equation}}

\newcommand{\halft}{\textstyle{\frac{1}{2}}}

\newcommand{\ee}{\textrm{e}}

\begin{document}
\numberwithin{equation}{section}

\title{Creep, recovery, and waves in a nonlinear fiber-reinforced
viscoelastic solid}

\author{M. Destrade,  G. Saccomandi}
\date{2007}

\maketitle

\begin{abstract}

We present a constitutive model capturing some of the experimentally 
observed features of soft biological tissues:
nonlinear viscoelasticity, nonlinear elastic anisotropy, 
and nonlinear viscous anisotropy. 
For this model we derive the equation governing rectilinear shear motion
in the plane of the fiber reinforcement;
it is a nonlinear partial differential equation for the shear strain. 
Specializing the equation to the quasistatic processes of creep 
and recovery, we find that usual (exponential-like) time growth 
and decay exist in general, but that for certain ranges of values 
for the material parameters and for the angle between the shearing 
direction and the fiber direction, some anomalous behaviors emerge.
These include persistence of a non-zero strain in the recovery experiment,
strain growth in recovery, strain decay in creep, disappearance of the 
solution after a finite time, and similar odd comportments.  
For the full dynamical equation of motion, we find kink (traveling wave) 
solutions which cannot reach their assigned asymptotic limit.

\end{abstract}

\section{Introduction}

Many biological, composite, and synthetic materials must
be modeled as fiber-reinforced nonlinearly elastic solids. Hence,
the anisotropy due to the presence of collagen fibers in many
biological materials has been studied extensively within the constitutive
context of fiber-reinforced materials by several authors (see for
example Humphrey (2002) and the  references therein.)
In nonlinear elasticity, the macroscopic response of an anisotropic material is given
in terms of a strain-energy function, which itself depends on a set of independent
deformation invariants. 
This formulation captures a great variety of phenomena related to
the behavior of fiber-reinforced materials such as \textit{inter alia}, 
the examination of fiber instabilities, using loss of ellipticity 
(see Merodio and Ogden 2002, 2003, and the references therein).

Generally speaking, a reinforcement is added to a given material
with the aim of avoiding a possible failure under operating
conditions. 
Therefore it is important to develop a detailed study showing
how  to introduce reinforcements in a material in order to 
control the possible development of a boundary layer structure. 
Our goal here is to provide a first step in this direction.
We make several simplifications and \emph{ad hoc} assumptions. 
First, we limit ourselves to the consideration of \textit{only one fiber direction} 
and second, we consider a \textit{one-dimensional motion} in the bulk of an infinite body. 
Here the motion is linearly polarized in a direction normal to the plane containing 
the direction of propagation and the direction of the fibers.  
We acknowledge that more complex anisotropies, geometries, and couplings arise 
in biomechanical applications. 
For instance, the mechanics of the aorta involves 
two families of parallel fibers, triaxial motions, and blood flow / arterial wall coupling. 
However we argue that some major characteristics of biological soft tissues are encompassed 
in the choices of transverse isotropy, of infinite extent, 
and of a motion governed by an ordinary differential equation.
Indeed the anisotropy due to the presence of one family of parallel fibers complicates 
the governing equations to an extent which is only marginally less than 
that due to the presence of two families of parallel fibers. 
Also, soft biological tissues are nearly incompressible and a (compressive) 
longitudinal wave is difficult to observe; it thus make sense to 
focus on transverse shear motions, which are useful in imaging technologies.
Our third assumption is that the elastic strain energy is the sum of an isotropic part and
an anisotropic part (called a \emph{reinforcing model}),
in order to model an isotropic base material 
augmented by a uniaxial reinforcement in the \emph{fiber direction}. 
Albeit strong, this constitutive assumption is now 
common and used by many authors (e.g. Triantafyllidis and
Abeyaratne 1983, Qiu and Pence 1997, Merodio and Ogden 2002).
Finally, we assume that the solid is viscoelastic and here we assume
not only Newtonian viscosity (proportional to the stretching tensor) 
but also fiber-oriented (anisotropic) viscosity.
That latter assumption is strong but can be removed from our calculations
by taking a constant to be zero. 
We believe that it might be useful in modeling the well-documented 
physiological effect of stretching training in sport medicine, which is 
that it affects the viscosity of tendon structures but not their elasticity
(Taylor et al., 1990; Kubo et al., 2002).

We divided the article into the following sections. 
Section 2 presents the constitutive model and the derivation of the 
equation governing the rectilinear shear motion. 
As expected, this equation is nonlinear in the shear strain:
it is a second-order partial differential equation, with cubic nonlinearity. 
To initiate its resolution, we first look at the quasistatic experiment of 
recovery in Section 3. 
Then we have a first-order ordinary differential equation, and we find 
that it can lead to unusual behaviors when certain conditions (strong anisotropy, 
large angle between the shearing direction and the fibers) are met. 
The same is true of the case of creep, treated in Section 4. 
Basically, it turns out that the nonlinearity introduces ranges of 
material parameters and angles for which an expected behavior 
-- say strain growth in creep -- can be turned on its head
-- and lead to strain decay with time in creep, say.
In the course of the investigation we develop synthetic tools 
of analysis which highlight the boundaries of these ranges. 
They also guide us for the resolution of the full dynamical equation
of motion, which we tackle in Section 5 for traveling wave solutions. 
Again the solution may behave in an unexpected way, provided the 
anisotropy is strong enough and the fibers are in compression. 
Finally, Section 6 recaps the results and puts them into a wider context.

\section{Basic equations} 

\subsection{The viscoelastic anisotropic model}

We describe the motion of a body  
by a relation $\mathbf{x}=\mathbf{x}(\mathbf{X},t)$, where $\mathbf{x}$
denotes the current coordinates of a point occupied by the particle of
coordinates $\mathbf{X}$ in the reference configuration at the time $t$.

We introduce $\mathbf{F} = \partial \mathbf{x} / \partial \mathbf{X}$, the deformation gradient,
and $\mathbf{C} = \mathbf{F}^\text{T} \mathbf{F}$, the
right Cauchy-Green strain tensor.
We focus on \textit{incompressible materials} for which all 
admissible deformations must be isochoric, or equivalently, 
for which the relation 
$\det \mathbf{F} = 1$ must hold at all times.

The body is reinforced with one family of parallel fibers.
Our first assumption is that the unit vector $\mathbf{a}_0$,
giving the fiber direction in the reference configuration, is independent 
of $\mathbf{X}$. 
The stretch along the fiber direction is $\sqrt{\mathbf{a}_{0}\mathbf{\cdot Ca}_{0}}
 = \sqrt{\mathbf{a \cdot a}}$ where $\mathbf{a} = \mathbf{F a}_0$.

We may now introduce the elastic part of our constitutive model.
We consider the so-called \emph{standard reinforcing model}, which is a quite simple
generalization to anisotropy of the neo-Hookean model 
(Triantafyllidis and Abeyaratne, 1983; Qiu and Pence, 1997).
For the standard reinforcing model, the strain-energy density is given by 
\begin{equation}
W = \frac{\mu}{2} \left[(I_1 - 3) + \gamma_0 (I_4 - 1)^2\right],
\quad
\text{where}
\quad
I_{1}= \text{tr }\mathbf{C},
\quad 
I_4 = \mathbf{a}_{0}\mathbf{\cdot Ca}_{0}= \mathbf{a \cdot a}.  \label{5}
\end{equation}
Here $\mu >0$ is the infinitesimal shear modulus of the 
isotropic neo-Hookean matrix, $\gamma _{0}>0$ is the \textit{elastic anisotropy parameter},
and the invariant $I_4$ measures the squared stretch in the fiber direction.
Mechanical tests show that the neo-Hookean strain energy function 
$\mu (I_1 - 3)/2$ fits uni-axial data rather well for arteries (Gundiah et al., 2006), 
while the anisotropic term  $\gamma_0 (I_4 - 1)^2$ is adequate to describe a reinforced material 
which penalizes deformation in the fiber direction (Merodio and Ogden, 2003).

The spatial velocity gradient $\mathbf{L} (\mathbf{X}, t)$
associated with a motion is defined as $\mathbf{L} = \text{grad } \mathbf{v}$,
where $\mathbf{v} = \partial \mathbf{x}/\partial t$ is the velocity, 
and the stretching tensor $\mathbf{D}$ is defined as 
$\mathbf{D} = \frac{1}{2}(\mathbf{L}+\mathbf{L}^{\text{T}})$.
For incompressible materials, $\text{tr } \mathbf{D} = 0$ at all times.
Newtonian viscous fluids possess a constitutive term in the form 
$2\nu \mathbf{D}$, where $\nu$ is a constant. 
For our special solid, we modulate the Newtonian viscosity with an anisotropic term, 
by replacing $\nu$ with $\nu[1 + \gamma_1 (I_4 -1)]$, where 
$\gamma_1>0$ is the \textit{viscous anisotropy parameter}. 
We show in the course of the paper that this simple choice 
of anisotropic viscosity captures the essential 
characteristics of attenuation in soft biological fibrous tissues. 
According to Baldwin et al. (2006), ultrasonic measurements of freshly excised myocardium show that 
``the attenuation coefficient was found to increase as a function of frequency in an approximately linear manner 
and to increase monotonically as a function of angle of insonification from a minimum perpendicular to 
a maximum parallel relative to the direction of the myofibers.''  

We are now ready to give the complete Cauchy stress tensor of our viscoelastic, 
transversally isotropic material as
\begin{equation}
\mathbf{T}  =  -p\mathbf{I} + \mu [\mathbf{B} 
                 + \gamma_0 (I_4 - 1)\mathbf{a \otimes a}] 
                   + 2\nu [1 + \gamma_1 (I_4 - 1)] \mathbf{D},
\label{T}
\end{equation}
where the $p$ is the yet indeterminate Lagrange multiplier introduced by the
incompressibility constraint, and $\mathbf{B} = \mathbf{F}\mathbf{F}^\text{T}$
is the left Cauchy-Green tensor.

\subsection{Shear motion}

We take a fixed, orthonormal triad of vectors ($\mathbf{i}$, $\mathbf{j}$, $\mathbf{k}$),
and call $X$, $Y$, $Z$ the reference coordinates; 
hence $\mathbf{X} = X\mathbf{i} + Y\mathbf{j} + Z\mathbf{k}$.
The triad is such that the unit vector in the fiber direction lies in the $XY$ plane;
hence,
\begin{equation}
\mathbf{a}_{0}=\cos \theta \mathbf{i} + \sin \theta \mathbf{j},
\label{rs1}
\end{equation}
(say) where $\theta \in [0,\pi]$ is the angle between the $X$-axis and the fibers. 

\begin{figure}
\centering \subfigure{\epsfig{figure=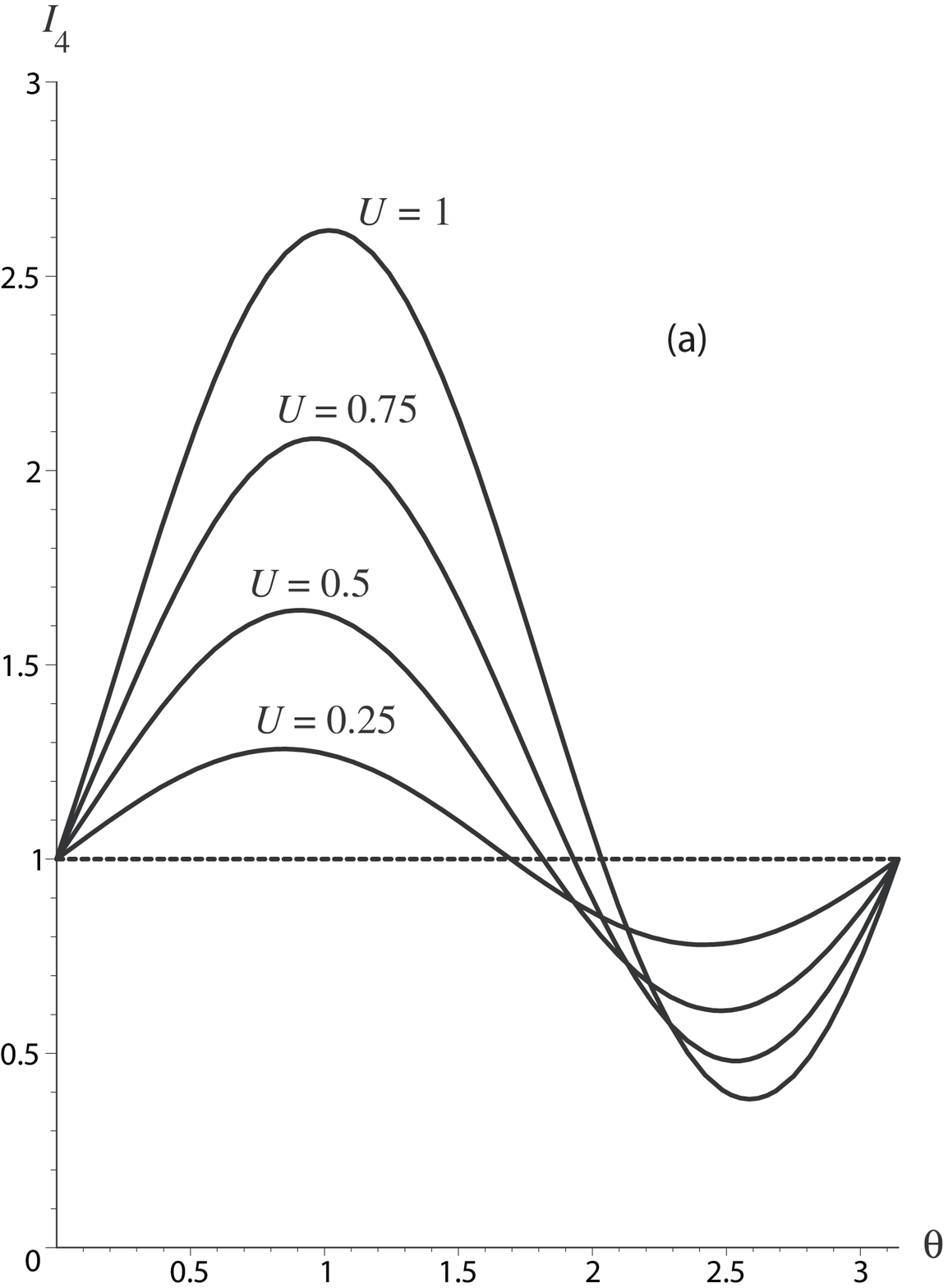,
width=.45\textwidth, height=.5\textwidth}}
  \quad
     \subfigure{\epsfig{figure=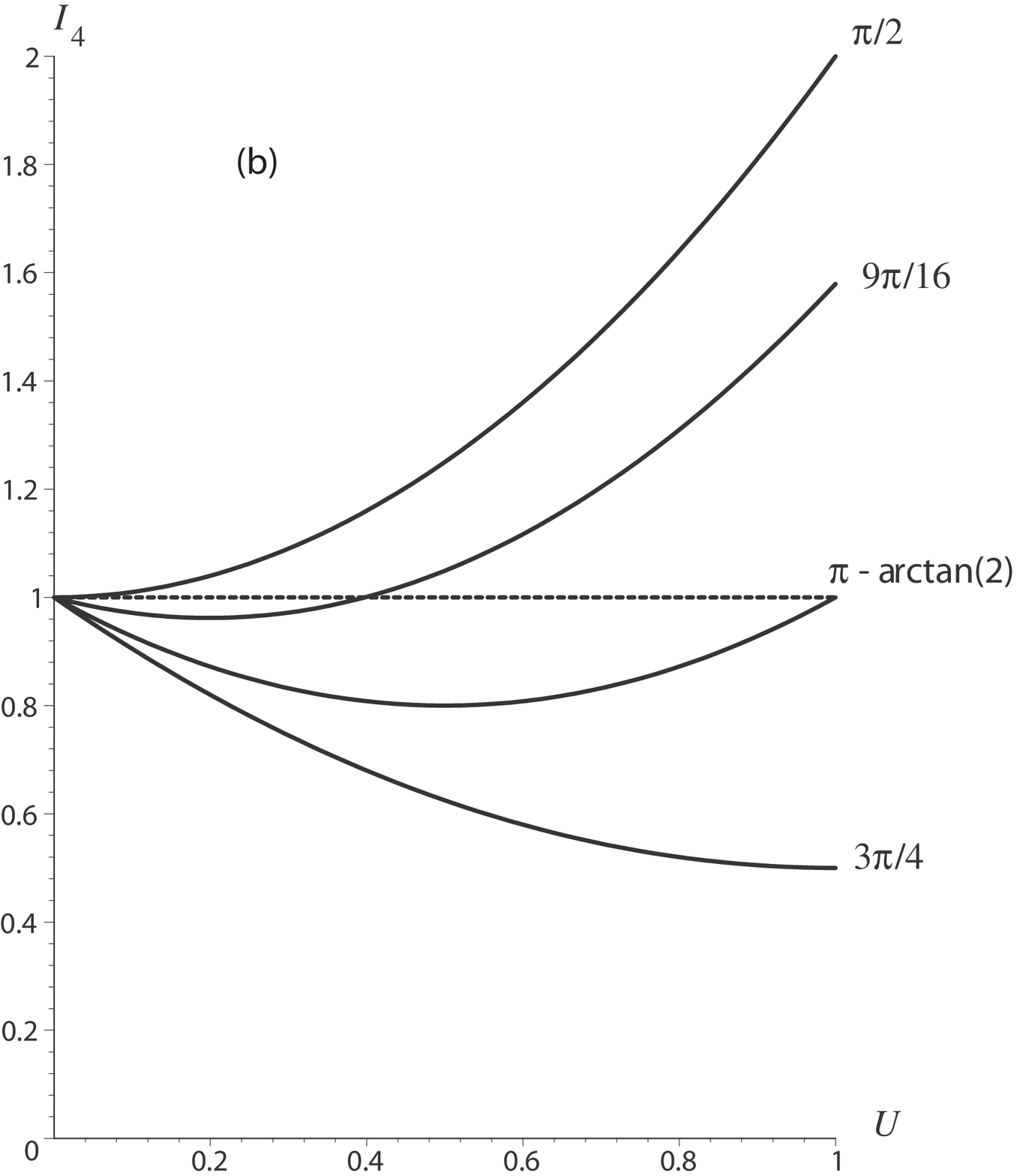,
width=.45\textwidth, height=.5\textwidth}}
 \caption{Variations of the squared stretch in the fiber direction: 
 (a) with the angle and (b) with the shear.
 When $I_4 > 1$, the fibers are in extension; when $I_4 < 1$, they are
  in compression.}
  \label{fig_1}
\end{figure}

We then consider the \textit{rectilinear shearing motion}, 
\begin{equation}
x= X + u(Y,t), \quad y=Y, \quad z=Z,  \label{rs2}
\end{equation}
where the anti-plane displacement $u$ is real and finite. 
Then the components of the gradient of deformation $\mathbf{F}$ 
and of its inverse are given by 
\begin{equation}
\mathbf{F} = \begin{bmatrix} 
                1 & U & 0 \\ 
                0 & 1 & 0 \\ 
                0 & 0 & 1
\end{bmatrix}, 
\qquad 
 \mathbf{F}^{-1} = \begin{bmatrix} 
1 & -U & 0 \\ 
0 & 1 & 0 \\ 
0 & 0 & 1
\end{bmatrix},  \label{rs3}
\end{equation}
where $U = \partial u / \partial Y$ is the \textit{amount of shear}.
The left and right Cauchy-Green tensors are thus 
\begin{equation}
\mathbf{B} = 
 \begin{bmatrix}
   U^2 + 1 & U & 0 \\ 
   U       & 1 & 0 \\ 
   0       & 0 & 1
\end{bmatrix},
\qquad 
\mathbf{C} = 
 \begin{bmatrix}
   1 & U & 0 \\ 
   U & U^2 + 1 & 0 \\ 
   0 & 0 & 1
 \end{bmatrix},  \label{rs5}
\end{equation}
respectively, 
from which the expressions of the invariants $I_1$ and $I_4$ follow,
\begin{equation}
I_1 = 3 + U^2, 
\quad 
I_4 = 1 + U \sin 2\theta + U^2 \sin^2 \theta.  \label{rs7}
\end{equation}
Figure \ref{fig_1}a shows the variations of $I_4$ with $\theta$ for several values of $U$ 
between 0 and 1. 
When $I_4 > 1$ the fibers are in extension, and when $I_4 < 1$ they are in compression;
the figure shows that this latter behavior occurs in a smaller and smaller angular range,
but is more and more pronounced,  as the amount of shear is increased.
Conversely, Figure \ref{fig_1}b shows the variations of $I_4$ with $U$ for several values of $\theta$;
when $0 < \theta < \pi/2$, the fibers are always in extension and when 
$\pi - \tan^{-1}(2) = 2.034 < \theta < \pi$, they are always in compression for $0 \leq U \leq 1$.
We refer to the paper by Qiu and Pence (1997) for similar figures and closely 
related discussions.

In the deformed configuration, we find that 
$\mathbf{a} = (\cos \theta + U \sin \theta)\mathbf{i} 
+ \sin \theta \mathbf{j}$. 
The remaining tensors required to compute the Cauchy stress tensor \eqref{T}
are
\begin{equation}
\mathbf{a \otimes a} =  
\begin{bmatrix}
 (\cos \theta + U \sin \theta)^2 &(\cos \theta + U \sin \theta) \sin \theta  & 0 \\ 
 (\cos \theta + U \sin \theta) \sin \theta  & \sin^2 \theta  & 0 \\ 
 0 & 0 & 0
\end{bmatrix},  \label{rs6}
\end{equation}
and
\begin{equation}
\mathbf{D} = \frac{1}{2} 
\begin{bmatrix}
0   & U_t & 0 \\ 
U_t & 0   & 0 \\ 
0   & 0   & 0
\end{bmatrix},
\end{equation}
so that the non-zero components of $\mathbf{T}$ are $T_{33}  =  -p+ \mu$
and
\begin{align}
& T_{11}  =  -p + \mu(1+U^2) + \mu\gamma_0 (I_4-1)(\cos \theta + U \sin \theta)^2, 
\notag \\ 
& T_{22}  =  -p + \mu +\mu \gamma_0 (I_4 - 1)\sin^2 \theta, 
\notag \\ 
& T_{12}  =  \mu U + \mu \gamma_0 (I_4-1)(\cos \theta + U\sin \theta) \sin \theta
                + \nu [1 + \gamma_1 (I_4 - 1)]U_t.
\label{rs8}
\end{align}

Now the equations of motion $\text{div } \mathbf{T} = \rho \mathbf{x}_{tt}$ reduce to the
two scalar equations: $-p_x + T_{12,y} = \rho u_{tt}$ and  $-p_y + T_{22,y} = \rho u_{tt}$.
Differentiating the former with respect to $y$ and the latter with respect to $x$, and eliminating 
$p_{xy}$, we arrive at a single governing equation for the rectilinear shear motion,
\begin{multline}
 \rho U_{tt}  = \mu U_{yy} +  \mu \gamma_0 \sin^2 \theta
    \left[U(2\cos \theta + U\sin \theta)(\cos\theta + U \sin \theta) \right]_{yy}
  \\ 
 +  \nu U_{tyy} + \nu \gamma_1 \sin\theta \left[U U_t  (2\cos \theta + U\sin \theta)\right]_{yy}.
\label{rs9}
\end{multline}
Using the scalings $\tilde{t} = \mu t/\nu$ and $\tilde{y} = y/L$ 
(where $L$ is a characteristic length to be specified later on a 
case-by-case basis), we write this equation in dimensionless
form as
\begin{multline}
 \varepsilon U_{\tilde{t}\tilde{t}}  = U_{\tilde{y}\tilde{y}} +  \gamma_0 \sin^2 \theta
    \left[U(2\cos \theta + U\sin \theta)(\cos\theta + U \sin \theta) \right]_{\tilde{y}\tilde{y}}
  \\ 
 +  U_{\tilde{t}\tilde{y}\tilde{y}} 
  +  \gamma_1 \sin\theta \left[U U_t  (2\cos \theta + U\sin \theta)\right]_{\tilde{y}\tilde{y}},
\label{nonDimension}
\end{multline}
where $\varepsilon =\rho \mu L^2 / \nu^2$. 
This is the main equation of our study. 
For convenience we drop the tildes in the remainder of the paper.
We also introduce the following functions,
\begin{align} \label{fg}
& f(\gamma_0, U,  \theta) =1+ \gamma_0 \sin^2 \theta(2\cos \theta + U\sin \theta)(\cos\theta + U \sin \theta), 
\notag \\
& g (\gamma_1, U, \theta)= 1 +  \gamma_1 U \sin\theta (2\cos \theta + U\sin \theta),
\end{align}
so that \eqref{nonDimension} is now
\begin{equation}
 \varepsilon U_{tt}  = 
    \left[U f(\gamma_0, U,  \theta) 
  +   U_t g (\gamma_1, U, \theta) \right]_{yy}.
\label{nonDimensionFG}
\end{equation}

\section{Nonlinear anisotropic recovery}

Our first investigation is placed in the quasistatic approximation, where
we study the influence of elastic anisotropy and viscous anisotropy 
on the classic experiment of viscous \textit{recovery}.
We imagine that the material is sheared and that at $t=0$ the shear stress is 
removed: $T_{12}(0) = 0$. 
Here the characteristic length $L$ is the displacement at $t=0$ 
from which the material will relax to the unstressed state.

In the quasistatic case, we neglect the inertia term of \eqref{nonDimension}
and may thus integrate it twice to give the following 
first-order ordinary differential equation, 
\begin{equation}
 U f(\gamma_0, U,  \theta) + U_t g (\gamma_1, U, \theta) =0.
\label{recovery}
\end{equation}
Here we take the constants of integration to be zero, according to the context of 
recovery, as explained above.
We then solve the equation as
\be \label{fgEqn}
\int \dfrac{g(\gamma_1, U, \theta)}
            {U f(\gamma_0, U, \theta)}
 \text{d}U = -t + \text{const.},
\en  
where the constant is computed so that $U(0) = 1$.

When $\theta = 0$, the fibers are not active with respect to the 
deformation and we recover the classical result of isotropic viscoelastic 
recovery: $U(t) =  \ee^{-t}$.

\begin{figure}
\centering \subfigure{\epsfig{figure=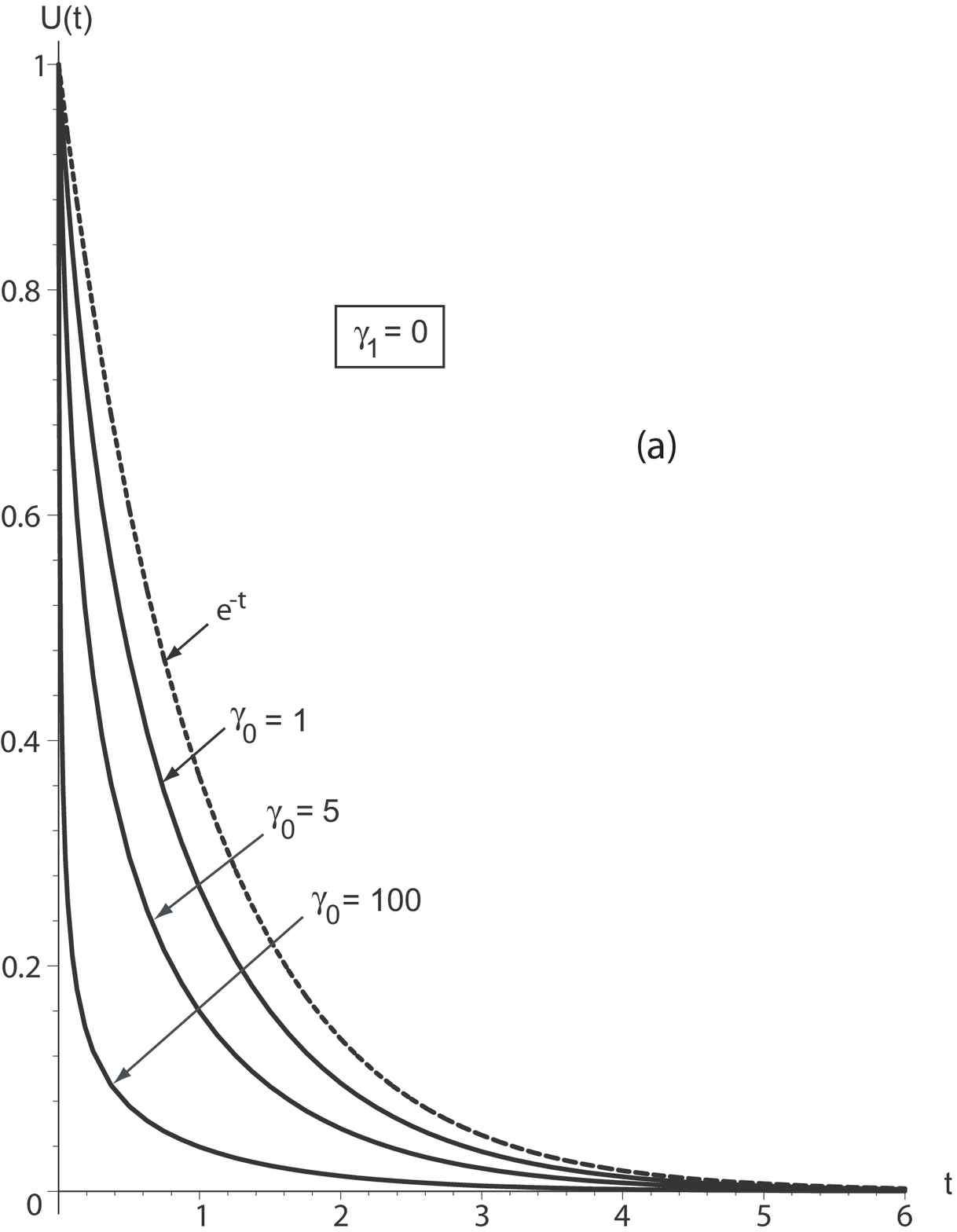,
width=.45\textwidth, height=.5\textwidth}}
  \quad
     \subfigure{\epsfig{figure=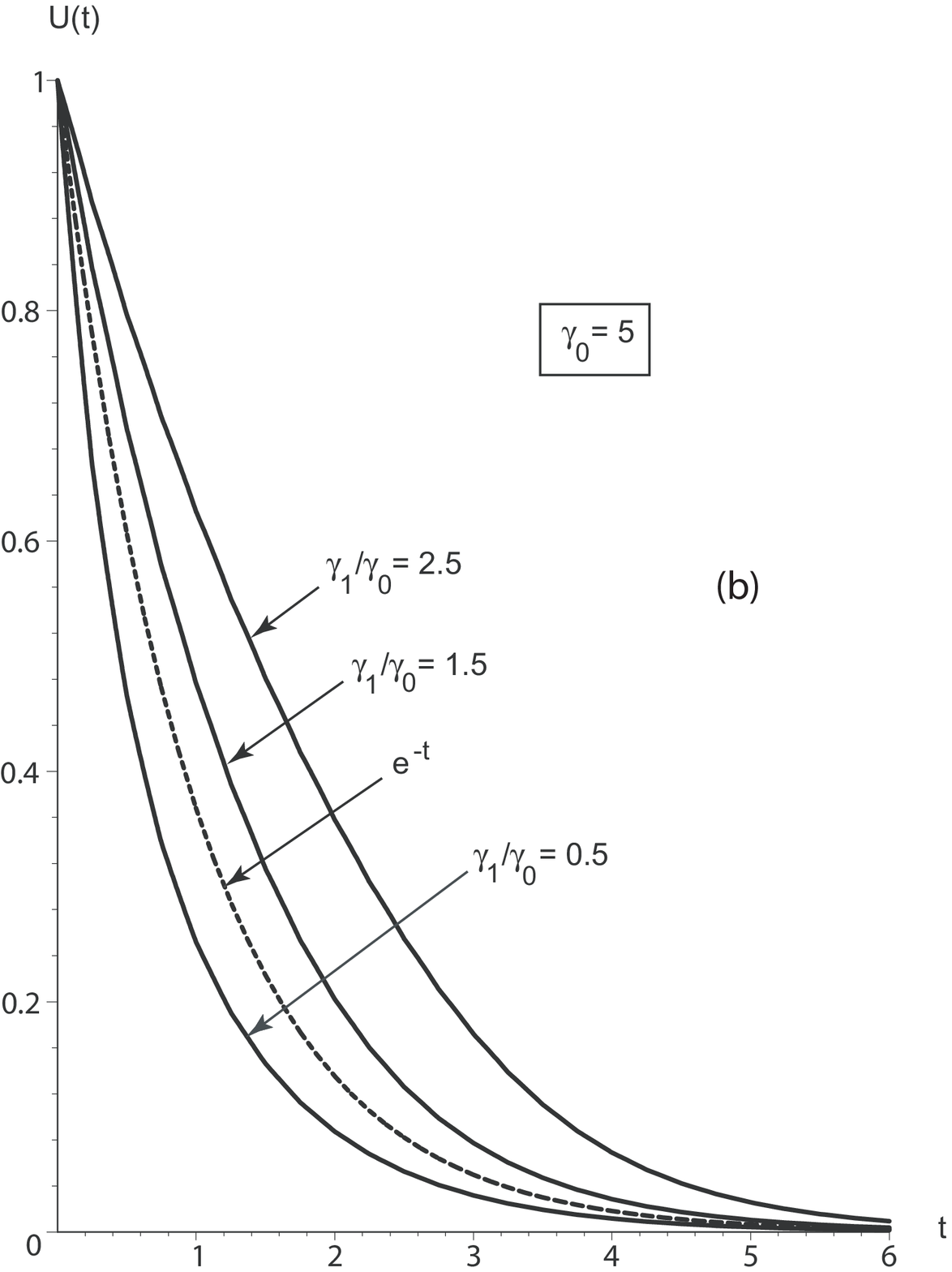,
width=.45\textwidth, height=.5\textwidth}}
 \caption{Time recovery function when $\theta =  \pi/2$: (a) $\gamma_1= 0$
 and $\gamma_0 = 1$, $5$, $100$;
 (b)   $\gamma_0= 5$
 and $\gamma_1 = 0.5$, $1.5$, $2.5$.
The recovery function for an isotropic solid is also plotted (dotted curve).}
\label{fig_2}
\end{figure}

When $\theta = \pi/2$, the anisotropic effects are at their strongest. 
In that case the integral above has a compact expression and we find:
\be
U \left[ \dfrac{1+\gamma_0 U^2}{1+\gamma_0} \right]^{\frac{1}{2}\left(\frac{\gamma_1}{\gamma_0}-1\right)}
  = \ee^{-t}.
\en
We now take $\gamma_1 =0 $ (no anisotropic viscosity) and 
$\gamma_0 = 1, 5, 100$ (recall that the fibers are inextensible in the 
limit $\gamma_0 \rightarrow \infty$).
Figure \ref{fig_2}a shows that as the anisotropic effect becomes more pronounced, 
the recovery is quicker; in other words, the influence of 
elasticity becomes stronger as $\gamma_0$ increases. 
Then we fix $\gamma_0$ at 5 for instance, and look at the 
role played by the anisotropic viscosity, by taking 
in turn $\gamma_1/\gamma_0 = 0.5$, $1.5$, 2.5.
We find in Figure \ref{fig_2}b that, as expected, the viscous recovery is slower 
as $\gamma_1$ increases.

\begin{figure}
 \centering 
  \epsfig{figure=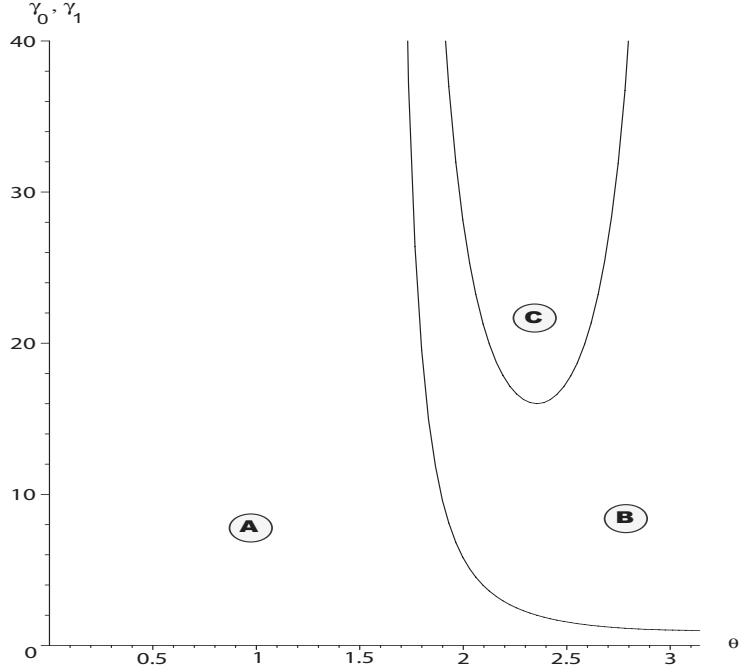, height=.65\textwidth, width=.7\textwidth}
\caption{Recovery: regions where the sign of $U_t$ may change.}
\label{fig_3}
\end{figure}

When $\theta \ne 0$, $\theta \ne \pi/2$, other behaviors arise, 
which call for a detailed analysis.
In particular, the exponential, or near-exponential, decay toward 
zero as $t \rightarrow \infty$ is not necessarily ensured, 
especially when the anisotropic effects are strong and the fibers are oriented 
at a large angle $\theta > \pi/2$. 
Clearly, $U_t = 0$ when $f=0$, according to \eqref{fgEqn}.
Also, $U_t<0$ when $f$ and $g$ are of the same sign and $U_t>0$
when $f$ and $g$ are of  opposite signs. 
These two functions are quadratic in $U$.
If they have no real roots in $U$, then they are both of the positive sign, and 
$U_t < 0$ (this is clearly the case in the region $0 < \theta < \pi/2$.)
If they have real roots, then they may change sign and $U$ might be an \emph{increasing}
function of $t$. 
This happens for $f$ and for $g$ when $\pi/2 < \theta < \pi$ and 
\be
\gamma_0 \geq \dfrac{4}{\cos^2\theta \sin^2 \theta}, \quad 
\gamma_1 \geq \dfrac{1}{\cos^2\theta},
\en
respectively.
In Figure \ref{fig_3}, the region C corresponds to the first inequality, where the delimiting 
curve has a vertical asymptote at $\theta = \pi/2$, 
a vertical asymptote at $\theta = \pi$, and a minimum at $\theta = \pi/4$, $\gamma_0=16$;
we recall that Qiu and Pence (1997) showed that when $\gamma_0 > 16$, ``simple shear at
certain fiber orientations involves negative shear stress in the shearing direction for
certain positive shears.''
The regions B and C correspond to the second inequality, where the delimiting 
curve has a vertical asymptote at $\theta = \pi/2$ and an  
horizontal asymptote at $\gamma_1 = 1$.

\subsection{Weak anisotropy}

First, we take both $\gamma_0$ and $\gamma_1$ in region A. 
This is the simplest case because $f$ and $g$ are then both positive and so 
$U_t$ is always negative (damped recovery).
We took several representative examples in this region 
(say $\theta = \pi/4$, $\gamma_0 = 20$, $\gamma_1 = 1$) and checked,
through integration and implicit plotting, that the graphs are indeed of the same
nature as those in Figure \ref{fig_2}.

\subsection{Strong elastic anisotropy}

\begin{figure}
\centering
\epsfig{figure=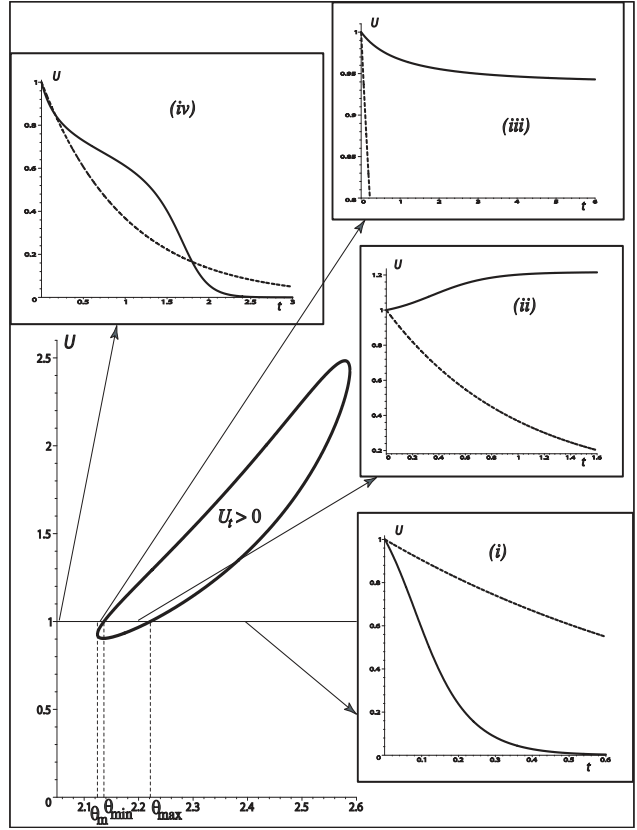,
width=.6\textwidth, height=.8\textwidth}
 \caption{Types of time recovery functions for $\gamma_0= 20$,
 $\gamma_1 = 1$ (strong elastic anisotropy). 
 The amount of shear $U$ starts at $1$ for $t=0$. 
 Outside the oval shape, $U_t < 0$ and $U$ decreases 
 as in ($i$) and ($iv$): decay toward zero; and in 
  ($iii$): decay toward a value $>0$. 
  Inside the oval shape, $U_t > 0$ and $U$ increases 
 as in ($ii$): growth toward a value $>1$.
The recovery function $\ee^{-t}$ for an isotropic solid is also plotted (dotted curves).}
\label{fig_4}
\end{figure}

Second, we take $\gamma_1$ in region A, by fixing it at $\gamma_1 = 1$, say.
In that region, $g>0$ always and thus the sign of $U_t$ is the opposite of the sign 
of $f$. 
Then we take $\gamma_0 = 20$, which is above the minimum of region C. 
In Figure \ref{fig_4}, we plotted the locus for the values of $U$ as functions of $\theta$ such that 
$f(20, U, \theta) = 0$. 
Outside the resulting oval shape, $f>0$, and inside, $f<0$.
We also plotted the line $U=1$, which intersects the oval at 
$\theta_\text{min} = 2.136$ and $\theta_\text{max} = 2.221$. 
Recall that $U(0)=1$.

When $\theta > \theta_\text{max}$, $U(t)$ starts at 1, and 
decreases because $f>0$ so that $U_t<0$;
as $U$ decreases toward 0, $U_t$ tends to zero according to \eqref{fgEqn}$_1$, 
but takes  an infinite time to do so, according to \eqref{fgEqn}$_2$;
hence $U=0$ is a horizontal asymptote and the recovery is ``classical'', see plot (i) 
in Figure \ref{fig_4}, traced at $\theta = 2.4$ (notice however that the recovery is not exponential 
because the second derivative of $U$ clearly changes sign as $t$ increases, 
in contrast with $\ee^{-t}$, traced in dotted line.)

When $\theta_\text{min} < \theta < \theta_\text{max}$, $U(t)$ starts at 1 and then 
grows until it hits the upper side of the oval, taking an infinite time to do so;
then this upper bound gives a horizontal asymptotic value, \emph{above the initial value}, 
see plot (ii) in Figure \ref{fig_4}, traced at $\theta = 2.2$. 

When $\theta_\text{m} < \theta < \theta_\text{min}$, where $\theta_\text{m} = 2.124$ is 
the angle at which the oval plot has a vertical tangent, $U(t)$ starts at 1 and then 
decreases until it hits the upper side of the oval, below 1 but above 0;
then this lower bound gives a horizontal asymptotic value, \emph{above zero}, 
see plot (iii) in Figure \ref{fig_4}, traced at $\theta = 2.125$.  

Finally, when $\theta < \theta_\text{m}$, $U(t)$ starts at 1 and then 
decreases until zero;
then this lower bound gives zero as a horizontal asymptotic value, 
see plot (iv) in Figure \ref{fig_4}, traced at $\theta = 2.05$. 
Notice that the second derivative changes signs three times as $t$ increases.  

\subsection{Strong viscous anisotropy}

\begin{figure}
\centering
\epsfig{figure=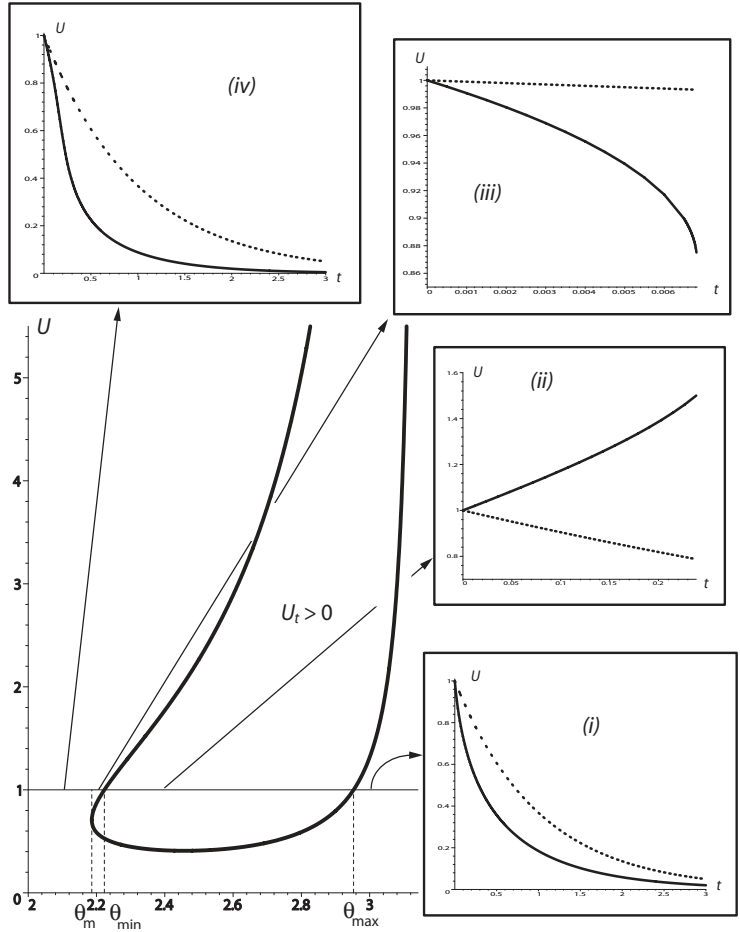,
width=.7\textwidth, height=.9\textwidth}
 \caption{Types of time recovery functions for $\gamma_0= 1$,
 $\gamma_1 = 2$ (strong viscous anisotropy). 
 The amount of shear $U$ starts at $1$ for $t=0$. 
 Outside the thick-line shape, $U_t < 0$ and $U$ decreases 
 toward zero as in ($i$) and ($iv$). 
  Inside the thick-line shape, $U_t > 0$ and $U$ increases 
 rapidly as in ($ii$), until it ceases to exist.
 There is also an angular region $\theta_\text{m} < \theta < \theta_\text{min}$
 where $U$ decreases 
 rapidly until it ceases to exist, see ($iii$).
The recovery function $\ee^{-t}$ for an isotropic solid is also plotted (dotted curves).}
\label{fig_5}
\end{figure}

Third, we take $\gamma_0$ outside the C region, by fixing it at $\gamma_0 = 1$, say.
In that region, $f>0$ always and thus the sign of $U_t$ is the opposite of 
the sign of $g$. 
Then we allow $\gamma_1$ to be in region B, and thus allow $g$ (and $U_t$) 
to change sign with increasing $\theta$, by taking $\gamma_1 = 3.0$ say. 
In Figure \ref{fig_5}, we plotted the values of $U$ as functions of $\theta$ such that 
$g(3, U, \theta) = 0$ and obtained the thick-line shape.  
Outside the shape, $g>0$, and inside, $g<0$.
We also plotted the horizontal line $U=1$, which intersects the shape at 
$\theta_\text{min} = 2.356$ and $\theta_\text{max} = 2.820$, 
and the vertical line $\theta = \theta_\text{m} = 2.186$, which is tangent to the shape.

Now when $\theta < \theta_\text{m}$ or $\theta > \theta_\text{max}$, 
$U(t)$ starts at 1 and decreases until zero;
as $U \rightarrow 0$, the denominator in the integral tends to 
zero, indicating that it takes an infinite time to do so; 
hence, zero is a horizontal asymptote in these cases.
To draw Figure \ref{fig_5}(i), we took  $\theta = 3.0$ and for Figure 
\ref{fig_5}(iv), we took $\theta = 2.1$; both graphs show a 
somewhat classical decay with time.

However, when  $\theta_\text{min} < \theta < \theta_\text{max}$, 
$U(t)$ starts at 1 and then \emph{grows} because $U_t > 0$ inside the 
thick line shape. 
Eventually $U$ hits the upper face of the shape, where $g = 0$;
then by \eqref{recovery}, either $U f =0$, or $U_t \rightarrow \infty$. 
Clearly, the first possibility is excluded because $U \ne 0$ when it is larger 
than 1, and $f \ne 0$ when $\gamma_0$ is outside the C region. 
It follows that $U$ grows and hits the upper face of the shape with a vertical asymptote 
after a finite time (and then stops because it cannot increase further since $U_t < 0$ outside the shape,
it cannot remain constant since $U_t \ne 0$ on the shape, and it cannot decrease since 
$U_t > 0$  inside the shape.)
Figure \ref{fig_5}(ii) shows such behavior for $U(t)$, traced at $\theta = 2.4$. 

Finally, when  $\theta_\text{m} < \theta < \theta_\text{min}$, 
$U(t)$ decays from 1 until it hits the shape from above after a finite time;
see Figure \ref{fig_5}(iii), traced at $\theta =2.2$. 
Notice how quickly the final value is reached, compared to the isotropic 
exponential recovery. 

\subsection{Strong elastic and viscous anisotropies}

In the case where both $\gamma_0$ and $\gamma_1$ are in the region C, 
any combination and overlaps of the thick curves presented in Figures \ref{fig_4} and 
\ref{fig_5} may arise. 
The tools presented in the two previous subsections are easily transposed to those 
possibilities. 
A special situation arises when the locus of $f = 0$ 
intersects the locus of $g = 0$;
then, the numerator and the denominator in \eqref{fgEqn} may 
have a common factor so that the integrand simplifies and a regular 
behavior may appear. 
This situation is however too special to warrant further investigation, and
we do not pursue this line of enquiry.

\section{Nonlinear anisotropic creep}

Our second investigation is again placed in the quasistatic approximation, where
we now study the influence of elastic anisotropy and viscous anisotropy 
on the classic experiment of viscous \emph{creep}.
As the resulting analysis is similar to that conducted for recovery, 
we simply outline the main results. 

We imagine that the material is sheared and that the shear stress is 
maintained: $T_{12}(\infty) \ne 0$. 
Here the characteristic length $L$ is an asymptotic value of the displacement.
We neglect the inertial term of \eqref{nonDimension}
and integrate it twice to give the following 
ordinary differential equation, 
\be
 U f(\gamma_0, U, \theta) + U_t g(\gamma_1, U, \theta)= \text{const.},
\label{creep}
\en
where we took the constant of the first integration to be zero, 
and the constant of the second integration to 
correspond to the applied (constant) shear stress, as is usual in the creep problem.
More specifically, this constant is taken so that $U(\infty) = 1$, and so is 
equal to $f(\gamma_0, 1, \theta)$; it follows that the equation above can be written as
\be \label{hgEqn}
h(\gamma_0, U, \theta) (U-1) + g(\gamma_1, U, \theta) U_t =0,
\en  
where $h$ is defined by 
\begin{align} \label{h}
h(\gamma_0, U, \theta) & = 
[U f(\gamma_0, U, \theta) - f(\gamma_0, 1, \theta)]/ (U-1)
\notag \\
&  = 
1 + \gamma_0 \sin^2 \theta [1 + \cos^2 \theta + (U+1) \sin \theta (U \sin \theta + 3 \cos \theta)].
\end{align}
We then solve the equation as
\begin{equation} \label{integrationCreep}
\int \dfrac{g(\gamma_1, U, \theta)}
            {(U-1) h(\gamma_0, U, \theta)}
 \text{d}U = -t + \text{const.},
\end{equation}
where the constant is computed so that $U(0) = 0$.
Hence  the equations governing creep are almost 
identical to those governing recovery, with the difference that 
$f$ is now replaced by $h$.

Here we are mostly concerned with the question of how, if at all, 
a state of shear can be reached such that once removed, the unusual 
recovery behaviors of the previous section emerge.
Thus we concentrate on strong anisotropic effects, with emphasis 
on strong elastic anisotropy (where the new function $h$ is involved).
We traced the regions where $g$ and $h$, and thus $U_t$, may 
change signs and found that the resulting graph is similar to that of 
Figure \ref{fig_3}, with the main difference that the minimum of region C is now located 
at $\theta = 3 \pi / 4$ and $\gamma_0 = 4$.
Thus unusual behavior in creep may occur at much lower levels of
elastic anisotropy than in recovery (where the minimum is at 
$\gamma_0 = 16$).
We recall that Qiu and Pence (1997) show that when  $\gamma_0 > 4$, 
``simple shear at certain fiber orientations involves a nonmonotonic
relation between the shear stress in the shearing direction and the amount of shear.''

\subsection{Strong elastic anisotropy}

\begin{figure}
\centering
\epsfig{figure=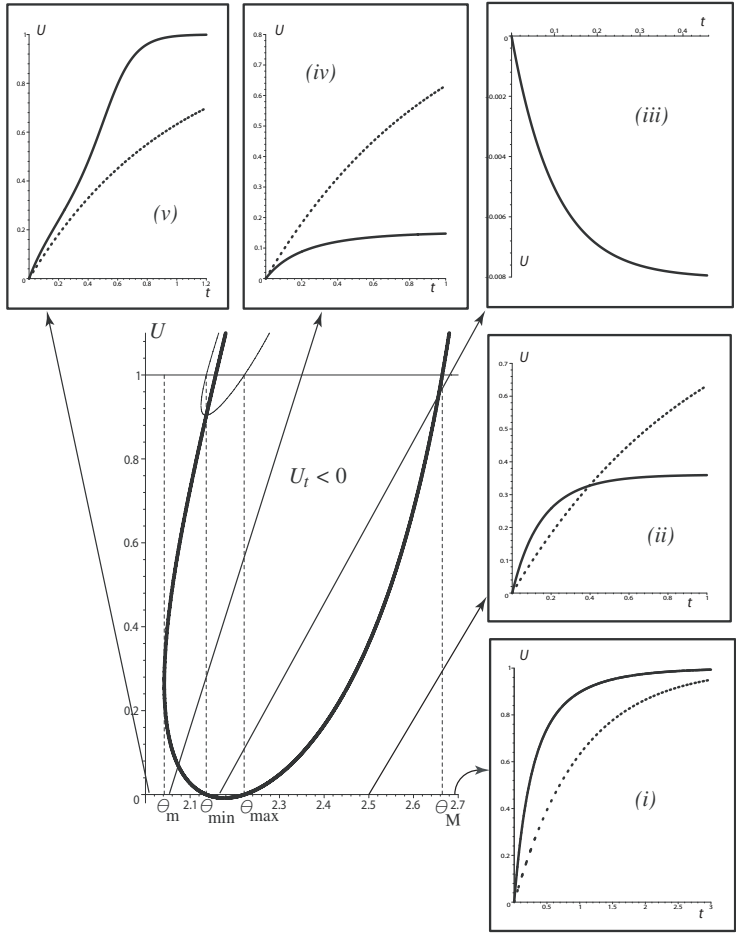,
width=.7\textwidth, height=.9\textwidth}
 \caption{Types of time creep functions for $\gamma_0= 20$,
 $\gamma_1 = 1$ (strong elastic anisotropy). 
 The amount of shear $U$ starts at 0 for $t=0$. 
 Outside the thick-line shape, $U_t > 0$ and $U$ increases 
 as in (i) and (v): growth toward $1$, and as in (ii) and (iv): 
 growth toward a value below $1$.
Inside the thick-line shape, $U$ decreases 
 as in (iii): decay toward a negative value.
The creep function $1 - \ee^{-t}$ for an isotropic solid is also plotted (dotted curves).}
\label{fig_6}
\end{figure}

We begin with the case where $h$ plays a major role, 
that is when $\gamma_0$ is greater than 4. 
For the purpose of direct comparison with the recovery problem, 
we take $\gamma_0 = 20$ and $\gamma_1 = 1$, as in Section 3.2. 
Figure \ref{fig_6} displays the curve where $h(20, U, \theta) = 0$. 
Outside the thick-line curve, $U_t > 0$, and inside, $U_t < 0$.
The curve intersects the line $U=0$ twice, at 
$\theta_\text{min} = 2.136$ and at $\theta_\text{max} = 2.221$. 
These are the values at which $f = 0$ intersects $U=1$ in Section 3.2
(see thin-line shape),
because by \eqref{h}, 
$h(20, 0, \theta_\text{min}) = f(20, 1, \theta_\text{min}) = 0$, 
and similarly $h(20, 0, \theta_\text{max}) = f(20, 1, \theta_\text{max}) = 0$.
We also display the vertical lines $\theta = \theta_M = 2.664$, where $h = 0$ intersects 
$U = 1$,  and  $\theta = \theta_m  = 2.042$, where $h = 0$ has a vertical tangent. 
Recall that for creep, $U(0)=0$.

When $\theta > \theta_\text{M}$, $U(t)$ starts at 0 and 
grows toward 1;
then $U_t$ tends to zero according to \eqref{hgEqn}, 
but takes  an infinite time to do so;
hence $U = 1$ is a horizontal asymptote and the creep is ``classical'', see plot (i) 
in Figure \ref{fig_6}, traced at $\theta = 2.7$ (the exponential creep function of isotropic 
visco-elasticity ($1 - \ee^{-t}$) is shown in dotted line.)

When $\theta_\text{max} < \theta < \theta_\text{M}$ or 
when $\theta_\text{m} < \theta < \theta_\text{min}$, $U(t)$ starts at 0 and then 
grows until it hits the oval shape, taking an infinite time to do so;
then this upper bound gives a horizontal asymptotic value, \emph{below 1}, 
see plot (ii) in Figure \ref{fig_6}, traced at $\theta = 2.5$, and 
plot (iv), traced at $\theta = 2.05$. 

When $\theta_\text{min} < \theta < \theta_\text{max}$, $U(t)$ starts at 0 inside the oval 
shape and thus it \emph{decreases} until it hits the lower side of the shape, taking an infinite time to do so;
then this lower bound gives a horizontal asymptotic value, \emph{below 0}, 
see plot (iii) in Figure \ref{fig_6}, traced at $\theta = 2.17$. 

Finally, when $\theta < \theta_\text{m}$, $U(t)$ can again grow toward 1,
see plot (v) in Figure \ref{fig_6}, traced at $\theta = 2.0$. 
Notice however that the concavity of the curve changes as $t$ increases.  

\subsection{Strong viscous anisotropy}

Here we remark that the function governing the strength 
of the viscous anisotropy, namely $g$, is the same for creep 
as it is for recovery. 
Thus, the region where $U_t$ might change sign because of strong 
viscous anistropy is the region B of Figure \ref{fig_2}. 
Also, the locus of points where $g=0$ is typically displayed by the 
thick-line shape of Figure \ref{fig_5} and 
because $g(\gamma_1, 0, \theta) = 1 >0$ always, this curve never crosses the abscissa $U=0$.
It follows that there is only one situation where viscous anisotropy 
leads to anomalous creep, when $\theta_\text{min} < \theta < \theta_\text{max}$;
then, $U(t)$ starts at zero, and grows toward the thick-line shape, which it 
reaches after a finite time with a vertical asymptote.

\subsection{Pre-stretch and nonlinear anisotropic creep}

\begin{figure}
\centering
\epsfig{figure=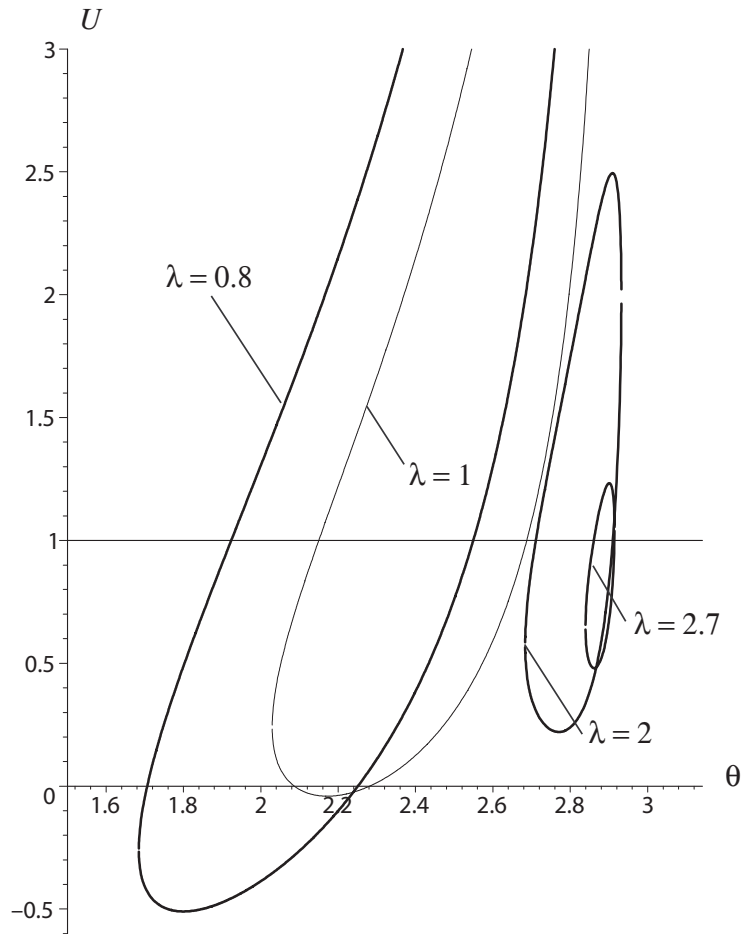,
width=.7\textwidth, height=.9\textwidth}
 \caption{Effect of pre-stretch on time creep functions for $\gamma_0= 30$,
 $\gamma_1 = 1$ (strong elastic anisotropy). 
 The amount of shear $U$ starts at 0 for $t=0$. 
 Inside the thick-line shapes, $U$ decreases; this situation may arise 
 when the solid is not pre-stretched ($\lambda = 1$) or when
 it is compressed ($\lambda = 0.8$). 
 Outside the thick-line shapes, $U_t > 0$ and $U$ increases; 
  this situation may arise 
  when the solid is in extension ($\lambda  = 2$, $\lambda =2.7$).}
 \label{fig_7}
\end{figure}

Here we show how anomalous creep can be avoided (amplified) by stretching (compressing)
the solid prior to the shear. 
Hence, instead of \eqref{rs2}, we consider the motion
\begin{equation}
x= \lambda^{-\halft} X + \lambda u(Y,t), \quad y = \lambda Y, 
\quad z= \lambda^{-\halft}Z,  \label{prestretch}
\end{equation}
The following decomposition of the associated deformation gradient 
shows that the solid is stretched by a ratio $\lambda$ in the $Y$
direction,
\be
\mathbf{F} = \mathbf{F_2 F_1},
\quad \text{where} \quad \mathbf{F_2} = \begin{bmatrix} 1 & U & 0 \\
		                         0 & 1 & 0 \\
                             0 & 0 & 1
             \end{bmatrix},
	   \quad
\mathbf{F_1} =             \begin{bmatrix} \lambda^{-\halft} & 0 & 0 \\
		                         0 & \lambda & 0 \\
                             0 & 0 & \lambda^{-\halft}
             \end{bmatrix}.
\en
(Note that $\mathbf{F_2 F_1} \ne \mathbf{F_1 F_2}$.) 
The kinematic quantities of Section 2.2 are modified accordingly.
In particular,
\be
I_4 = \lambda^{-1} \cos^2\theta  + \lambda^{2} \sin^2\theta             
 + U \lambda^{\halft} \sin 2\theta + U^2 \lambda^{2} \sin^2\theta.    
\en
The end result is that the differential equation governing creep is 
changed from \eqref{hgEqn} to
\be \label{hgEqnPreStretch}
h^\lambda(\gamma_0, U, \theta) (U-1) + g^\lambda(\gamma_1, U, \theta) U_t =0,
\en  
where $h^\lambda$ and $g^\lambda$ are defined by 
\begin{align} \label{hLambda}
& h^\lambda(\gamma_0, U, \theta)  = 
\lambda^2 \left\{1 + \gamma_0 \sin^2 \theta 
 [2 \lambda^2 \sin^2 \theta  + 3 \lambda^{-1}\cos^2 \theta - 1 
 \right.
 \notag \\
& \phantom{ h^\lambda(\gamma_0, U, \theta)  = 
\lambda^2 \left\{1 + \gamma_0 \sin^2 \theta 
 [2 \right.}\left. 
  + (U+1) \sin \theta (U \sin \theta + 3 \cos \theta)]\right\},
 \notag \\
& g^\lambda(\gamma_1, U, \theta)  = 
 1 + \gamma_1 (\lambda^{-1} \cos^2 \theta +
  \lambda^2 \sin^2 \theta - 1 
  + U \lambda^{\halft} \sin 2\theta + U^2 \lambda^2 \sin^2 \theta).
\end{align}
Figure \ref{fig_7} shows the loci of $h^\lambda = 0$ in the case of a strong elastic 
anisotropy ($\gamma_0= 30$,
$\gamma_1 =$ 1), for several values of $\lambda$.
The figure clearly shows that the pre-stretch $\lambda$ can be used to control 
the shape of these curves: if the solid is put in compression first, 
and sheared for creep next, then the region of potential anomalous 
creep is increased; if it is put under tension, then the area of the region 
rapidly decreases and eventually dissapears altogether.  

\section{Nonlinear traveling waves}

So far we have looked at how the presence of elastic and viscous fibers 
affects some quasi-static processes. 
Typically, creep and recovery connect one state of constant shear (initial) 
to another (final).
Now we examine another class of solutions connecting two constant states of
shear, this time \emph{dynamically}, by looking for traveling wave (kink) 
solutions. 

The mathematical theory of one-dimensional transverse traveling waves in
isotropic viscoelastic materials with a Kelvin-Voigt type of constitutive
equation is well grounded,  see for example Nishihara (1995) for a clear
and complete mathematical approach, or Jordan and Puri (2005) for a specific and
explicit example. 
A traveling wave is a solution to the equations of motion in the form
\begin{equation}
U(Y,t) = U(\xi ), \qquad  \xi = Y-ct,  \label{tw1}
\end{equation}
where $c$ is the constant speed; also, $U$ is such that 
\begin{equation}  \label{tw2}
\lim_{\xi \rightarrow -\infty } U(\xi ) = U_{L}, 
\qquad 
\lim_{\xi \rightarrow \infty}U (\xi )=U_{R}, 
\end{equation}
where $U_{L}$ and $U_{R}$ are distinct constants. 
In what follows, we focus on the case where $U_{L}=0$, $U_{R}=1$. 
This case is general up to a rigid translation. 
Here we take the displacement corresponding to $U_{R}$ as the characteristic length $L$.

Substituting \eqref{tw1} into  \eqref{fgEqn}, we obtain 
\begin{equation}
\varepsilon c^2 U'' = \left(U f -  c U' g\right)'',  \label{tw3}
\end{equation}
and then by integration,
\begin{equation}
c U' g = \left(f - \varepsilon c^2 \right)U + \text{const.} \label{tw4}
\end{equation}
By the requirement $U_L = 0$, the constant must be zero. 
By the requirement $U_R = 1$, we have  
\begin{equation} \label{f(1)}
f(\gamma_0, 1, \theta) = \varepsilon c^2.  
\end{equation}
This equation prompts three remarks.

\begin{figure}
\centering \subfigure{\epsfig{figure=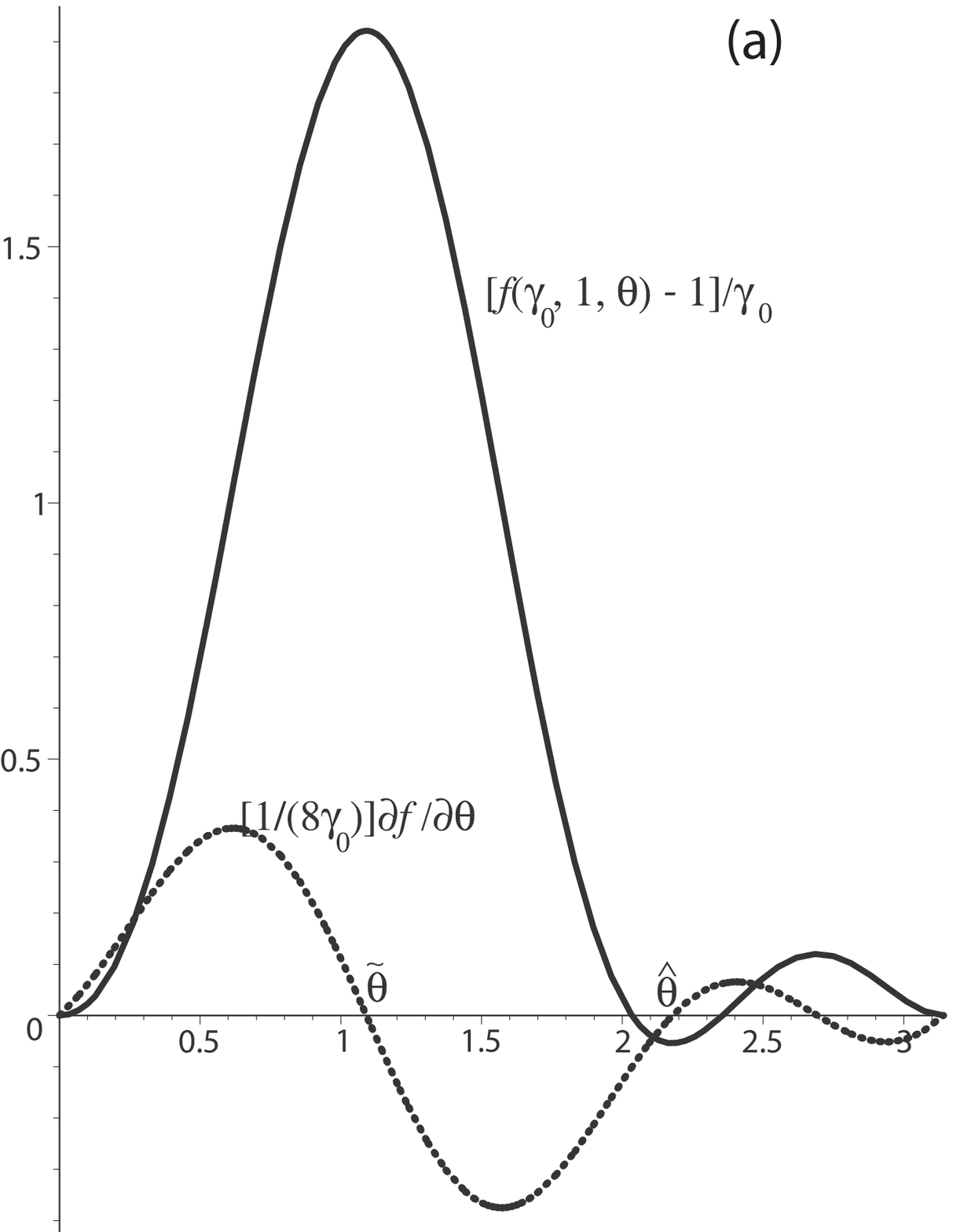,
width=.45\textwidth, height=.5\textwidth}}
  \quad
     \subfigure{\epsfig{figure=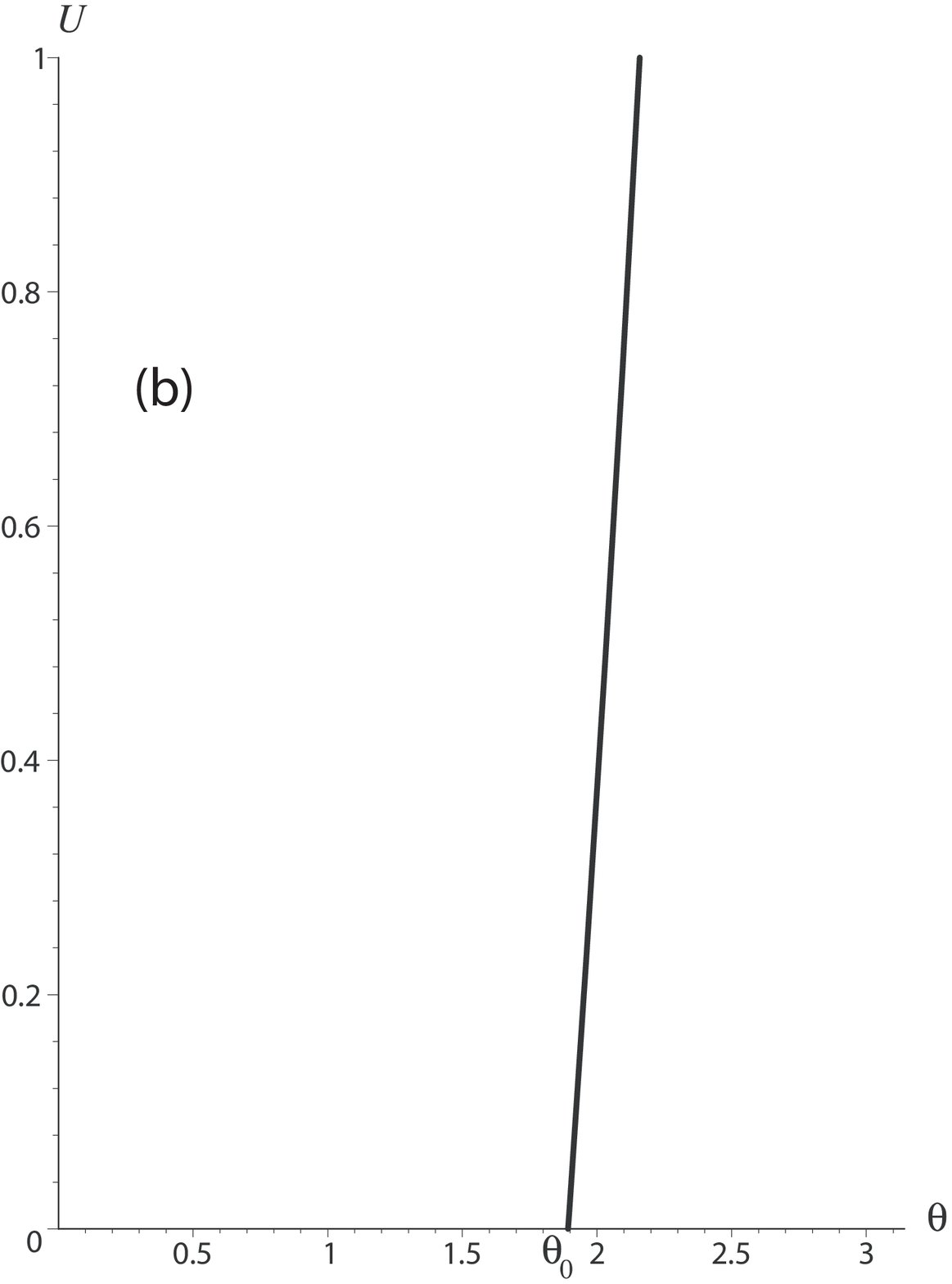,
width=.45\textwidth, height=.5\textwidth}}
 \caption{(a) Variations with $\theta$ of $[f(\gamma_0, 1,  \theta) - 1]/\gamma_0$ 
and of its derivative, 
showing an absolute minimum at $\hat{\theta} = 2.1777$. 
(b) Variations of $-3 / \tan \theta - 1$ with $\theta$, crossing the abscissa line at 
$\theta_0 = 1.8926$.}
\label{fig_8}
\end{figure}

First, we must ensure that $f(\gamma_0, 1, \theta) >0$.
Recall that according to \eqref{fg},
\be \label{f(1,theta)}
 f(\gamma_0, 1,  \theta) = 1 + \gamma_0 \sin^2 \theta(2\cos \theta + \sin \theta)(\cos\theta + \sin \theta),
\en
and so,
\be \label{Df(1,theta)}
 \partial f(\gamma_0, 1,  \theta) / \partial \theta = \gamma_0 \sin \theta(4\cos^3 \theta + 9\sin \theta \cos^2 \theta - 3\sin^3 \theta).
\en
In Figure \ref{fig_8}a we plotted the variations of $[f(\gamma_0, 1,  \theta) - 1]/\gamma_0$
with $\theta$, as well as those of its derivative with respect to $\theta$ (scaled to $1/8$).
Clearly, the function  \eqref{f(1,theta)}, viewed as a function of $\theta$, 
has an absolute minimum and an absolute maximum. 
The minimum is at $\hat{\theta}$, say, 
such that $\tan \hat{\theta}$ is that root of the cubic $4 + 9x - 3x^3 = 0$ 
corresponding to $\pi/2 < \hat{\theta} < \pi$;
numerically, $\hat{\theta} = 2.1777$. 
Then, solving $f(\gamma_0, 1, \hat{\theta}) = 0$ for $\gamma_0$, 
we find that  $f(\gamma_0, 1, \theta) >0$ when 
$0 < \gamma_0 < \hat{\gamma_0} = 18.490$;
and that when $\gamma_0 > \hat{\gamma_0}$, there appears a range 
for $\theta$ where  $f(\gamma_0, 1, \theta) >0$  is not insured.
Placing ourselves outside that possibility, we deduce from \eqref{f(1)}
that for a given $\gamma_0$ and a given $\theta$, the wave travels with 
speed
\begin{equation}
 c = \pm \sqrt{f(\gamma_0, 1, \theta) / \varepsilon}.  \label{tw6}
\end{equation}
This is of course expressed in the dimensionless variables of length$/L$ and 
time$\times \mu / \nu$.
Turning back, if required, to physical variables, we would find that the wave 
travels with the dimensional speed $\sqrt{\mu f(\gamma_0, 1, \theta) / \nu}$.

The second remark is that according to \eqref{f(1)} and \eqref{f(1,theta)}, the wave 
(when it exists) travels with maximum speed at the angle $\tilde{\theta}$, say, 
such that $\tan \tilde{\theta}$ is that root of the cubic $4 + 9x - 3x^3 = 0$  
corresponding to $0 < \tilde{\theta} < \pi/2$;
numerically, $\tilde{\theta} = 1.0910$.
Hence the directions of extremal speeds of propagation are always the same, whatever the values of 
the constitutive parameters $\mu$, $\gamma_0$, and $\gamma_1$ are. 
This observation indicates the way for an acoustic determination of the fiber orientation:
if an experimental measurement of the shear wave speed can be made in every direction of 
a fiber-reinforced viscoelastic nonlinear material, then the fibers are at an angle $\hat{\theta}$
from the direction of the slowest wave and at an angle $\tilde{\theta}$
from the direction of the fastest wave.
We recall that for waves in an \emph{isotropic} deformed neo-Hookean material, Ericksen (1953) 
found that the fastest waves propagate along the direction of greatest initial stretch. 

The third remark is that when \eqref{f(1)} holds, then 
\begin{equation}
f(\gamma_0, U, \theta) - \varepsilon c^2 
 = \gamma _{0} U (U-1) \sin^3 \theta \left[U \sin \theta 
               + 3 \cos \theta +  \sin \theta \right].  \label{tw7}
\end{equation}
Then the separation of variables, followed by integration of the first order
differential equation \eqref{tw4}, leads to 
\begin{equation}
\int \frac{g(\gamma_1, U, \theta)}
       {U(U-1) \sin^3 \theta [U \sin \theta 
               + 3 \cos \theta +  \sin \theta]}
           \text{d}U 
             = \frac{\gamma _{0}}{c} \xi + \text{const.},
    \label{tw8}
\end{equation}
where the constant of integration is arbitrary;
without loss of generality, we take it to be such that $U(0) = 1/2$.

\begin{figure}
\centering \subfigure{\epsfig{figure=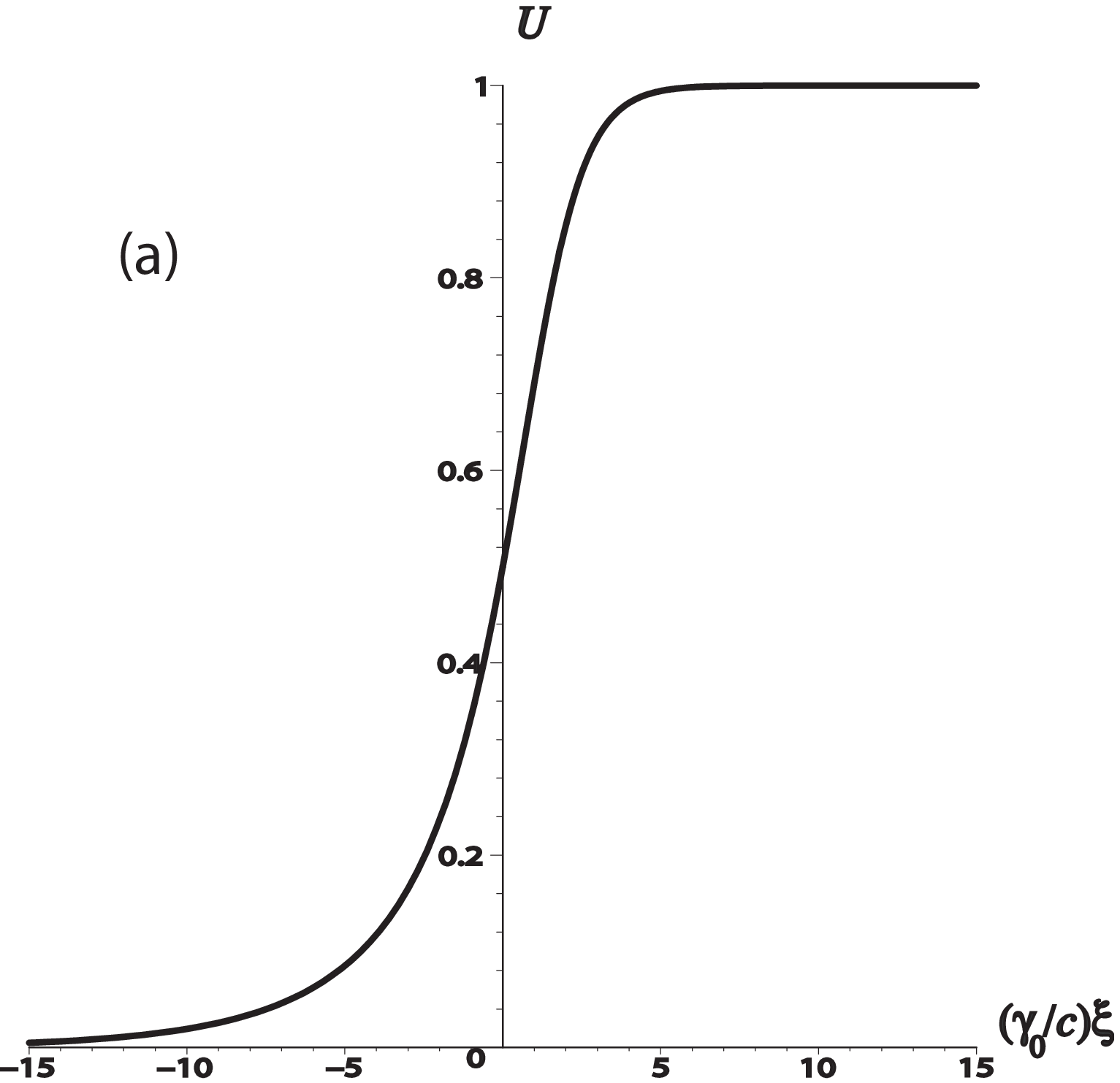,
width=.45\textwidth, height=.5\textwidth}}
  \quad
     \subfigure{\epsfig{figure=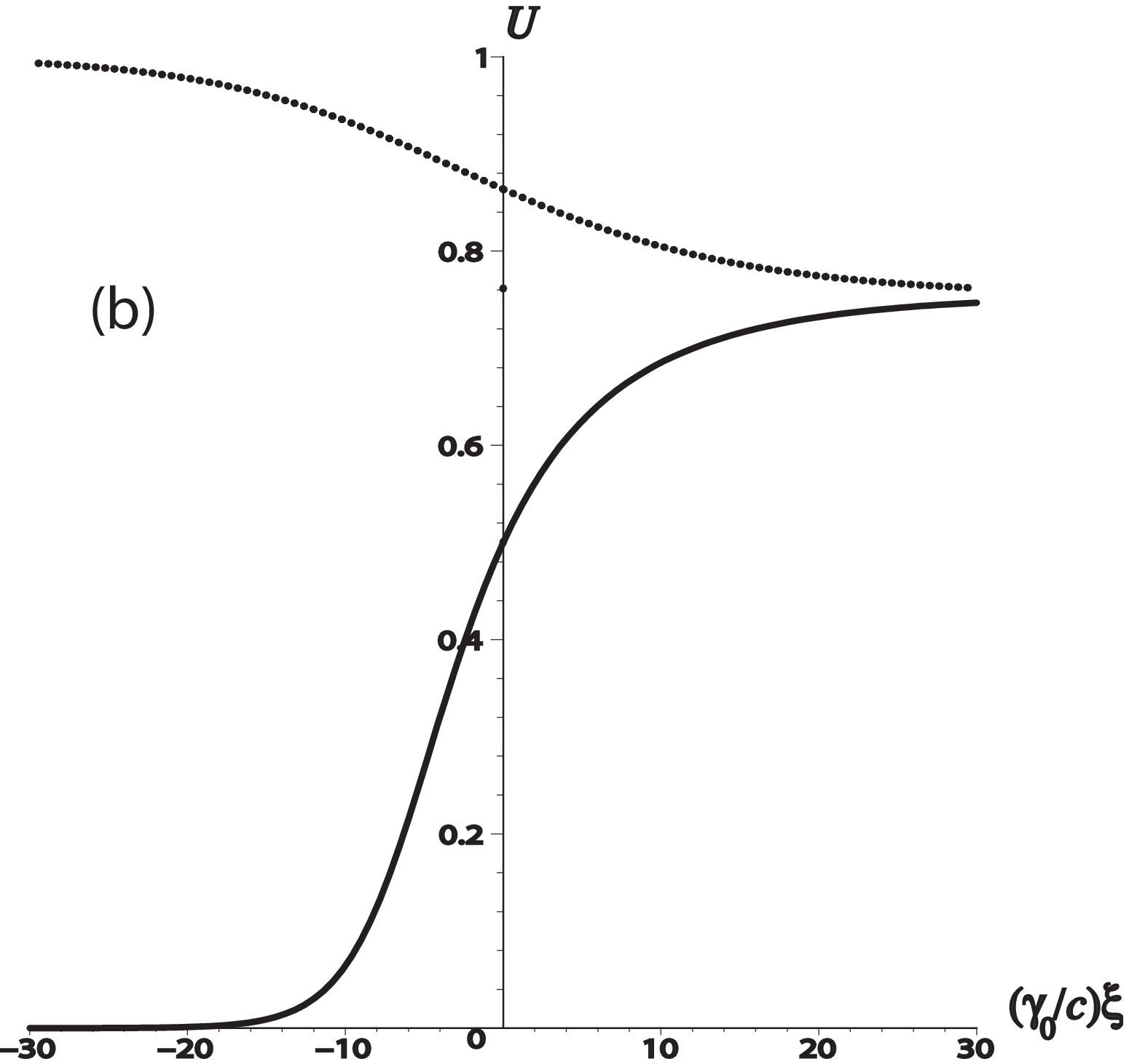,
width=.45\textwidth, height=.5\textwidth}}
 \caption{Traveling wave solution for anisotropic elasticity ($\gamma_1 = 0$); 
 (a) at $\theta = 1.8$,  
 (b) at $\theta = 2.1$.}
 \label{fig_9}
\end{figure}

Clearly, critical issues arise when either the numerator or the denominator change 
signs (because then $U'$ changes sign and it might not be possible 
to find a solution satisfying the requirements \eqref{tw2}).
We may take care of the numerator's sign by considering elastic anisotropy only
($\gamma_0 \ne 0$) and discarding viscous anisotropy ($\gamma_1 = 0$); 
then $g = 1$. 
For the denominator however, we note that $U \sin \theta + 3 \cos \theta + \sin \theta$ 
can change sign for certain ranges of $U$ and $\theta$. 
Figure \ref{fig_8}b shows the curve $U = -3 / \tan \theta - 1$; on its left side, 
the denominator is positive, on its right side, it is negative. 
Accordingly, the wave connects 0 to 1, see in Figure \ref{fig_9}a, or 
is unable to do so, see Figure \ref{fig_9}b. 
In that latter case, the wave front grows toward an asymptotic value 
which is less than 1; a second solution exist (dotted curve) with 1 
as an asymptotic value, but in the $\xi \rightarrow -\infty$ direction.

As a final remark, we note that when $\gamma_1$ is large enough to allow for 
the possibility that $g = 0$ (strong viscous anisotropy), then a ``singular 
barrier'' arises, see Pettet et al. (2000).

\section{Discussion}

In the course of this investigation on nonlinear anisotropic creep, 
recovery, and waves for fiber-reinforced nonlinear elastic materials, 
we unearthed  some complex mechanical responses. 
For some range of the constitutive parameters and for some angle ranges of 
the fiber arrangement, we saw that unusual and possibly aberrant behaviors can emerge. 

From a mathematical point of view, we gave a detailed explanation of the reasons for
these behaviors, by linking them  to the singularities of the determining equations for the 
amount of shear. 

From the mechanical point of view, we pointed out that  
non-standard behaviors always occur when the angle between the fiber family and
the direction of shear is such that the fibers are compressed, see Figure \ref{fig_1}. 
It has been widely demonstrated that several types of instabilities may 
develop in the case of fiber contraction, see the detailed studies 
by Triantafyllidis and Abeyaratne (1983), 
Qiu and Pence (1997), Merodio and Ogden (2002, 2003, 2005ab), or 
Fu and Freidin (2004).
For example, Merodio et al. (2007) recently investigated a 
non-homogeneous rectilinear shear static deformation for the 
standard reinforcing model \eqref{5} and found non-regular solutions 
(that is, deformations characterized by a discontinuous amount of shear) 
in fiber-contracted materials.

From a numerical point of view,
we recall a simple model, together with a simple class of solutions,
allows a step-by-step control of the simulations. 
It would indeed be hard to detect non-standard behaviors by relying solely 
on a complex numerical finite element method, and omitting to conduct a simple analytical 
methodology such as the one presented in this paper. 
For example, Holzapfel and Gasser (2001) present a detailed computational 
study of some viscoelastic fiber-reinforced nonlinear materials, 
but use values for the material parameters and for the angles which place 
their simulations outside the problematic ranges. 
Other studies are placed in the framework of linear models (even for 
polymeric materials, see Liu et al., 2006), which fail to 
capture non-standard behaviors.

From an experimental point of view, our results 
suggest some simple yet revealing protocols. 
In particular, it would be most valuable to investigate 
the existence and the persistence of asymptotic residual shear strains, 
sustained after the shear stress is removed, at levels not only below
the value at initial time but also above (as in  Section 3).  
So far we have only identified reports of experimental
results concerned with elastomeric materials reinforced with \emph{inextensible} fibers 
(and therefore with a ratio between the shear modulus of the bulk matrix and that 
of the fibers of several orders of magnitude), or concerned with moderate 
angles between the direction of shear and the fiber direction.

From a biomechanical point of view, the results have meaningful implications for 
biological soft tissues.
First, the model captures adequately the elastic and the viscous anisotropies of 
biological materials (Baldwin et al. 2006, Taylor et al. 1990). 
Second, although anomalous creep behaviors might preclude anomalous recovery 
behaviors, it is still useful to study the latter, because they might 
nonetheless arise in vivo following a stress-driven fiber orientation remodeling
(Hariton et al., 2006).
Third, the effect of the prestretch on non-standard behaviors is significant theoretically 
(Section 4.3) as well as practically (in vivo experiments show that large static
prestretches of tendons reduces the risk of unexpected behaviors, see Kubo et al. (2002)).
Finally, the results of the traveling wave study (Section 5) 
may eventually lead to an acoustic (elastographic) determination of the fiber angle
in soft tissues, through an efficient, simple, and non-invasive investigation.
 
Obviously, our results must be improved and several directions are possible. 
Hence, two families of fibers have to be considered to give a 
better comparison with in vivo results for soft tissues.
Also, the more realistic models of fiber reinforcements (such as 
the one proposed by Horgan and Saccomandi (2005) and by Gasser et al. (2006))
must be incorporated in the present study, to 
identify with a greater precision the range of parameters for which strange behaviors may occur.


\end{document}